\documentclass[aip,pop,reprint]{revtex4-1}
\usepackage{color}
\usepackage{graphicx}
\usepackage{amsmath}
\usepackage{amssymb}
\usepackage{cancel}
\usepackage{subfig}
\usepackage{bm}
\graphicspath{{./figs/}}

\renewcommand{\vec}[1]{\mathbf{#1}}

\newcommand{\ddt}[2]{\frac{\partial{#1}}{\partial{#2}}}

\newcommand{\comment}[1]{}
\newcommand{\ig}[2]{\includegraphics[width = #1]{#2}}

\begin{document}
\title{Using the maximum entropy distribution to describe electrons in reconnecting current sheets}
\author{Jonathan Ng}
\author{Ammar Hakim}
\author{A. Bhattacharjee}
\affiliation{Center for Heliophysics, Department of Astrophysical Sciences and Princeton Plasma Physics Laboratory, Princeton University, Princeton, NJ 08543, USA}
\date{\today}
\begin{abstract}
Particle distributions in weakly collisional environments such as the magnetosphere have been observed to show deviations from the Maxwellian distribution. These can often be reproduced in kinetic simulations, but fluid models, which are used in global simulations of the magnetosphere, do not necessarily capture any of this. We apply the maximum entropy fluid closure of Levermore, which leads to well posed moment equations, to reconstruct particle distributions from a kinetic simulation in a reconnection region. Our results show that without information other than the moments, the model can reproduce the general structure of the distributions but not all of the finer details. The advantages of the closure over the traditional Grad closure are also discussed.
\end{abstract}
\maketitle

\section{Introduction}

In weakly collisional plasma environments, particle distributions can become quite non-Maxwellian. This is supported by observational evidence in space plasmas, such as the solar wind, where measured proton and electrons can show distortions with anisotropy and heat flux \cite{marsch:2006}, the magnetopause, with recently observed ``crescent'' electron distributions in reconnecting regions \cite{burch:2016}, and near reconnection regions in the magnetotail \cite{oka:2016}, where anisotropic, flat-top and other complex distributions have been seen.

For magnetic reconnection -- a change in topology of the magnetic field lines in a plasma \cite{dungey:1953} -- in particular, kinetic simulations have been used to study these distributions and their origin in detail \cite{le:2009, bessho:2014, bessho:2017}. However, global simulations of magnetospheres, which include these reconnection regions, use fluid models such as magnetohydrodynamics (MHD). Though there have been efforts to extend fluid models to include aspects of the kinetic physics using higher moment equations \cite{yin:2001,wang:2015,miller:2016,ng:2017}, the closure of the moment equations is difficult, and the underlying distributions of these models do not represent the complex kinetic structure. 

This work focuses on understanding which aspects of the distribution function can be represented correctly by moment closures. To that end, we employ the principle of maximum entropy \cite{jaynes:1957}, which provides a method to determine a probability distribution given limited information. This has been used in many scientific fields including astrophysics \cite{richstone:1988}, biology \cite{yeo:2004} and natural language processing \cite{charniak:2000}. With regard to fluid closures, the maximum entropy fluid closure of Levermore \cite{levermore:1996} has the additional advantage of leading to well-posed fluid equations, which is not true for traditional methods such as the Grad method \cite{grad:1949}. 

The rest of paper is organised as follows: We first describe the traditional Chapman-Enskog and Grad approaches to the closure of moment equations, followed by the derivation of the maximum entropy closure of Ref.~\onlinecite{levermore:1996} in Section~\ref{sec:closures}. This closure is then used to reconstruct distribution functions close to the electron diffusion region of a kinetic simulation of magnetic reconnection in Section~\ref{sec:dists}. We conclude in Section~\ref{sec:conclusion} with a summary and discussion of the implications of our findings.

\section{Maximum Entropy Fluid Closure}
\label{sec:closures}

In this section we review the derivation of the maximum entropy closure of Levermore \cite{levermore:1996}. This was devised as a nonperturbative alternative to the usual Grad moment systems and ensures that the hierarchy of moment equations obtained is hyperbolic. 

We start with the Boltzmann equation for a single particle phase space distribution $f(\vec{x},\vec{v},t)$,
\begin{equation}
\ddt{f}{t} + \vec{v}\cdot\nabla f = \mathcal{C}(f).
\end{equation}
Here $f$ is the single-particle phase space density and $\mathcal{C}(f)$ is a collision operator, assumed to conserve number, momentum and energy. The evolution of fluid quantities is obtained by taking moments of this equation to get the hierarchy of fluid equations \cite{levermore:1996, grad:1949}.

\begin{equation}
\begin{split}
\ddt{n}{t} &+ \nabla\cdot(n \vec{v}) = 0 \\
\ddt{n m \vec{v}}{t} &+ \nabla \cdot (\vec{P} + n m \vec{v}\vec{v}) = 0 \\
&\vdots
\end{split}
\end{equation}
Here $n$ is the number density, $\vec{v}$ is the fluid velocity, $\vec{P}$ is the pressure tensor. 

More generally, one can write 
\begin{equation}
\ddt{\vec{U}}{t} + \nabla\cdot \vec{F} = \vec{S},
\end{equation}
where $\vec{U}$ is the vector of moments of the distribution function, $\vec{F}$ is the tensor of the associated fluxes and $\vec{S}$ the vector of source terms arising from moments of the collision operator. The problem of closure arises as the time evolution of each fluid quantity depends on the flux term, which contains higher velocity moments, so that there are always more moments than equations. It is then necessary to describe these unknowns in terms of other quantities (usually the lower moments) to close the equations. 

In the Chapman-Enskog approach \cite{chapman:1970}, the distribution function is described by small deviations from local equilibrium, and can be expanded in powers of the Knudsen number $\epsilon$, which describes the ratio of the mean free path to the gradient scale. 
\begin{equation}
\begin{split}
f_{CE}(\vec{x},\vec{v},t) &= f_M(\vec{x},\vec{v},t)(1 + \epsilon f^{(1)} + \epsilon^2 f^{(2)} + \dots),\\
f_M(\vec{x},\vec{v},t) &= \frac{m^{3/2}n(\vec{x},t)}{\left(2\pi T(\vec{x},t)\right)^{3/2}} \exp\left( - \frac{m |\vec{v} - \vec{u}(\vec{x},t)|^2}{2 T(\vec{x},t)} \right)
\end{split}
\end{equation}
The fluid equations can then be closed by solving for the distribution function using this expansion. Truncating the expansion at the zeroth order leads to the Euler equations, which are adiabatic and inviscid, while the first order expansion leads to the Navier-Stokes equations, where viscosity and heat flux are expressed in terms of velocity and temperature gradients respectively \cite{levermore:1996}. Beyond the Navier-Stokes equations, one obtains the Burnett and super-Burnett equations, which must be modified to prevent instability at short wavelength \cite{bobylev:2008}. 

In the Grad approach, the distribution function is expanded in Hermite polynomials about a local Maxwellian, and closure is achieved by truncating the Hermite expansion. This can be written (omitting the $\vec{x}$ and $t$ dependence for brevity) as \cite{grad:1949}
\begin{equation}
f_{Grad}(\vec{v}) = f_M(\vec{v}) \sum_{n=0}^\infty \frac{1}{n!}\alpha_i^{(n)}H_i^{(n)}\left(\frac{\vec{w}}{v_t}\right)
\end{equation}
where $\vec{w} = \vec{v} - \vec{u}(\vec{x},t)$ and $H^{(n)}_i$ are multivariate Hermite polynomials \cite{grad:1949hermite}. The coefficients $\alpha$ are determined from the moments of the distribution function. For clarity, the distribution function can also be written in terms of physical quantities. For example, the 20-moment model \cite{grad:1949}, which is used in the next section, is
\begin{equation}
f_{20}(\vec{v}) = f_M(\vec{v})\left(1 + \frac{\pi_{ij}}{2 p T} w_i w_j + \frac{q_{ijk}}{6 p T^2} w_i w_j w_k - \frac{q_i}{2 p T} w_i\right).
\end{equation}
Here Einstein summation convention is used, and $p$ is the scalar pressure, $\pi_{ij} = p_{ij}-p\delta_{ij}$ is the traceless part of the pressure tensor, and $q_{ijk}$ and $q_i$ are components of the heat flux tensor and vector respectively. Although moment equations can be derived using this closure, the Grad distribution functions can become negative in regions of phase space, and the moment equations can become ill-posed away from equilibrium \cite{levermore:1996,cai:2015}. In spite of these issues, the regularised thirteen moment system has had some success in describing rarefied gases \cite{struchtrup:2003,torrilhon:2004}, and is still the subject of study using non-Grad distribution functions \cite{torrilhon:2010,ottinger:2010,singh:2016}.

In contrast to the previous approaches, the maximum entropy approach introduced by Levermore \cite{levermore:1996} is non-perturbative and uses a distribution function 
\begin{equation}
f(\bm{\alpha},\vec{v}) = \exp (\vec{\bm{\alpha}}^\tau \vec{m}(\vec{v})), 
\label{eq:levermore}
\end{equation}
where $\vec{m}(\vec{v})$ a vector containing monomials up to a certain degree of the particle velocity components and $\vec{\bm{\alpha}}$ is a vector of closure coefficients \cite{levermore:1996}. 

As the distribution function must be finite as $v\rightarrow\infty$, the polynomial in the exponent must be of an even degree. For maximal degree 2, the 5 or 10-moment equations are obtained, with the density, momentum and either scalar or tensor pressure being evolved. At degree 4, the next two systems, which are considered in this paper, are the 14 and 21-moment equations, which retain the heat flux vector and tensor in a manner similar to the 13- and 20-moment Grad equations. The additional quantity is the $|v|^4$ moment, which is necessary to ensure the distribution remains finite as mentioned earlier. 

In this closure model, the free parameters are the coefficients $\vec{\bm{\alpha}}$. These are determined by a process which maximises entropy given the known moments of the distribution. Here we sketch the derivation \cite{levermore:1996}. 

Using the form of the distribution function given above, the moment equations can be rewritten as 
\begin{equation}
\ddt{\langle \vec{m}(\vec{v}) f(\bm{\alpha},\vec{v}) \rangle}{t} + \nabla\cdot \langle \vec{v}\vec{m}(\vec{v}) f(\bm{\alpha},\vec{v}) \rangle = \langle \vec{m}(\vec{v}) \mathcal{C}(f(\bm{\alpha},\vec{v}))\rangle,
\label{eq:levmoment}
\end{equation}
where the angled brackets indicate integration over velocity space. 

From here, we omit the $\vec{v}$ argument and let $h(\vec{\bm{\alpha}}) = \langle f(\vec{\bm{\alpha}})\rangle$ and $\vec{j}(\vec{\bm{\alpha}}) = \langle \vec{v} f(\vec{\bm{\alpha}})\rangle$. These are denoted as the density and flux potentials respectively in Ref.~\onlinecite{levermore:1996}. Here we can see that the $\bm{\alpha}$ derivatives of these quantities give the moment and flux terms in Eq.~\eqref{eq:levmoment}. Thus the moment equations can be written as 
\begin{equation}
\partial_t h_{\bm{\alpha}} + \nabla \cdot j_{\bm{\alpha}} = \vec{S}(\bm{\alpha})
\label{eq:denpot}
\end{equation}

Equation~\eqref{eq:denpot} can be rewritten as 
\begin{equation}
h_{\bm{\alpha}\bm{\alpha}}(\vec{\bm{\alpha}}) \partial_t \vec{\bm{\alpha}} + \vec{j}_{\bm{\alpha}\bm{\alpha}}(\vec{\bm{\alpha}})\cdot \nabla \vec{\bm{\alpha}} = \vec{S}(\bm{\alpha}).
\end{equation}
This ensures hyperbolicity as $h_{\bm{\alpha}\bm{\alpha}}(\vec{\bm{\alpha}})$ is positive definite and $j_{\bm{\alpha}\bm{\alpha}}(\vec{\bm{\alpha}})$ is symmetric. 

The closure coefficients $\bm{\alpha}$ are determined through the minimisation of the quantity $h(\bm{\alpha}) -  \langle \bm{\alpha}^\tau \vec{m} f(\bm{\alpha})\rangle$ with respect to $\bm{\alpha}$. This maximises the entropy $-\langle f \log f - f \rangle$ given the constraint of the known moments \cite{levermore:1996}. The remaining quantities in the moment equations can then be calculated using this distribution. 

For five and ten moments, there are closed form solutions, giving the isotropic Maxwellian $f_5(\vec{v}) = f_M(\vec{v})$, and the generalised Gaussian distribution 
\begin{equation}
f_{10}(\vec{v}) = \frac{n}{\sqrt{(2\pi)^3\det(\Theta)}} \exp\left(-\frac{1}{2} \left(\vec{v}-\vec{u}\right)^\tau\Theta^{-1}\left(\vec{v}-\vec{u}\right) \right)
\end{equation}
where $\Theta$ is a positive definite matrix. 

There are two main limitations of this technique. For moments with degree greater than $2$, there is no closed form solution and the coefficients must be calculated numerically using a minimisation process. It has also been shown that there can exist physically realisable states which cannot be described by Eq.~\eqref{eq:levermore}, though it is possible to guarantee realisability by slightly modifying the distribution \cite{junk:1998, mcdonald:2013, groth:2009}. Fluid simulations of the 14-moment (in 2-D) and 35-moment (in 1-D) systems have been performed and show good agreement with kinetic methods, but there are still ongoing efforts to improve their efficiency \cite{mcdonald:2013interp,boone:2016,schaerer:2017}.

\section{Application to reconnection}
\label{sec:dists}

When magnetic reconnection occurs in weakly collisional environments such as the magnetosphere, the particle distributions can become highly non-Maxwellian as has been shown in observations \cite{chen:2008,burch:2016} and kinetic simulations \cite{egedal:2013,bessho:2014,le:2017}. We have chosen this system to show how even large deviations from the Maxwellian can be described by the Levermore model. This section demonstrates the reconstruction of the particle distributions at various points in the reconnection region using the 10-, 14- and 21-moment maximum entropy models. These models are of particular relevance as the electrons exhibit strong pressure anisotropy close to reconnection regions \cite{egedal:2013}, and the divergence of the pressure tensor balances the reconnection electric field in the diffusion region \cite{hesse:2008}. While the role of the heat flux is not as well understood, fluid models simulations of reconnection have been shown to be sensitive to its precise form \cite{hesse:2004,ng:2015,ng:2017,allmann:2018}.

\begin{figure}
\ig{3.375in}{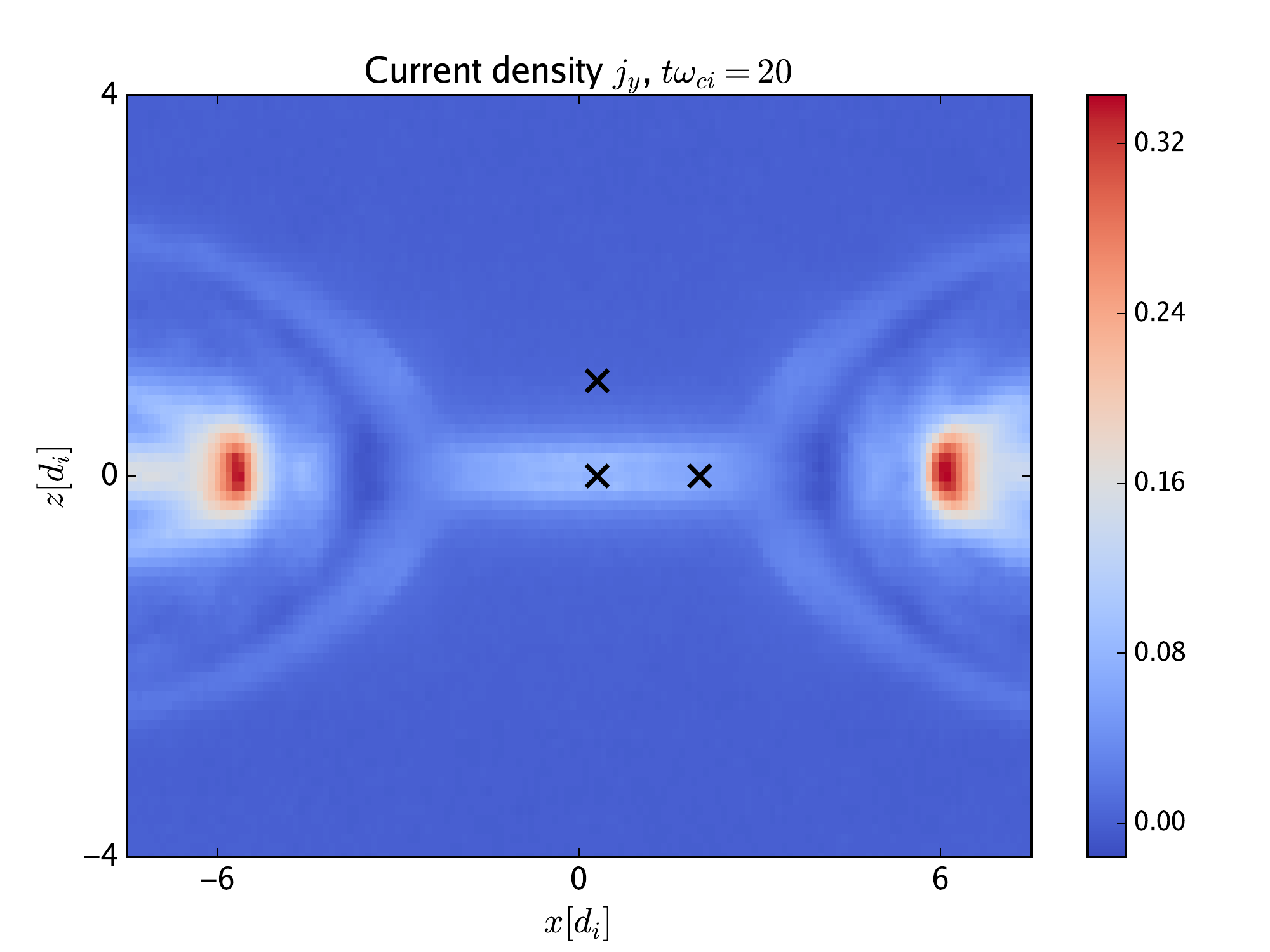}
\caption{Structure of the current sheet at $t\omega_{ci} = 20$. The $y$-axis is into the page. Crosses mark positions where the upstream, x-point and exhaust particle distributions are studied. }
\label{fig:sheet}
\end{figure}

To reconstruct these distributions, we perform a kinetic simulation of a Harris sheet \cite{harris:1962} using the particle-in-cell code PSC \cite{germaschewski:2016}. Parameters are similar to the antiparallel GEM scenario \cite{birn:2001}, with $m_i/m_e = 25$, $\omega_{pe}/\omega_{ce} = 2$, $T_i/T_e = 5$. Reconnection is initiated by a small magnetic field perturbation. The moments are taken directly from the simulation data, with reference particle distributions extracted from boxes with side $2.5 d_e$ around the points marked in Fig.~\ref{fig:sheet} at $t\omega_{ci} = 20$. These correspond to the inflow, x-point and exhaust regions and there are approximately 10000 particles in each sampling region. 

From the known moments, the coefficient vector $\bm{\alpha}$ is determined iteratively using a Newton method. We note that some of the electron distributions close to the x-line contain relativistic electrons, so the distributions below are shown in $\vec{p}$-space. However, the closure model uses a non-relativistic approximation, so that $\vec{p}= m_e\vec{v}$ and the form of the maximum-entropy distribution in the previous section remains the same in these calculations. 

\begin{figure}
\ig{3.375in}{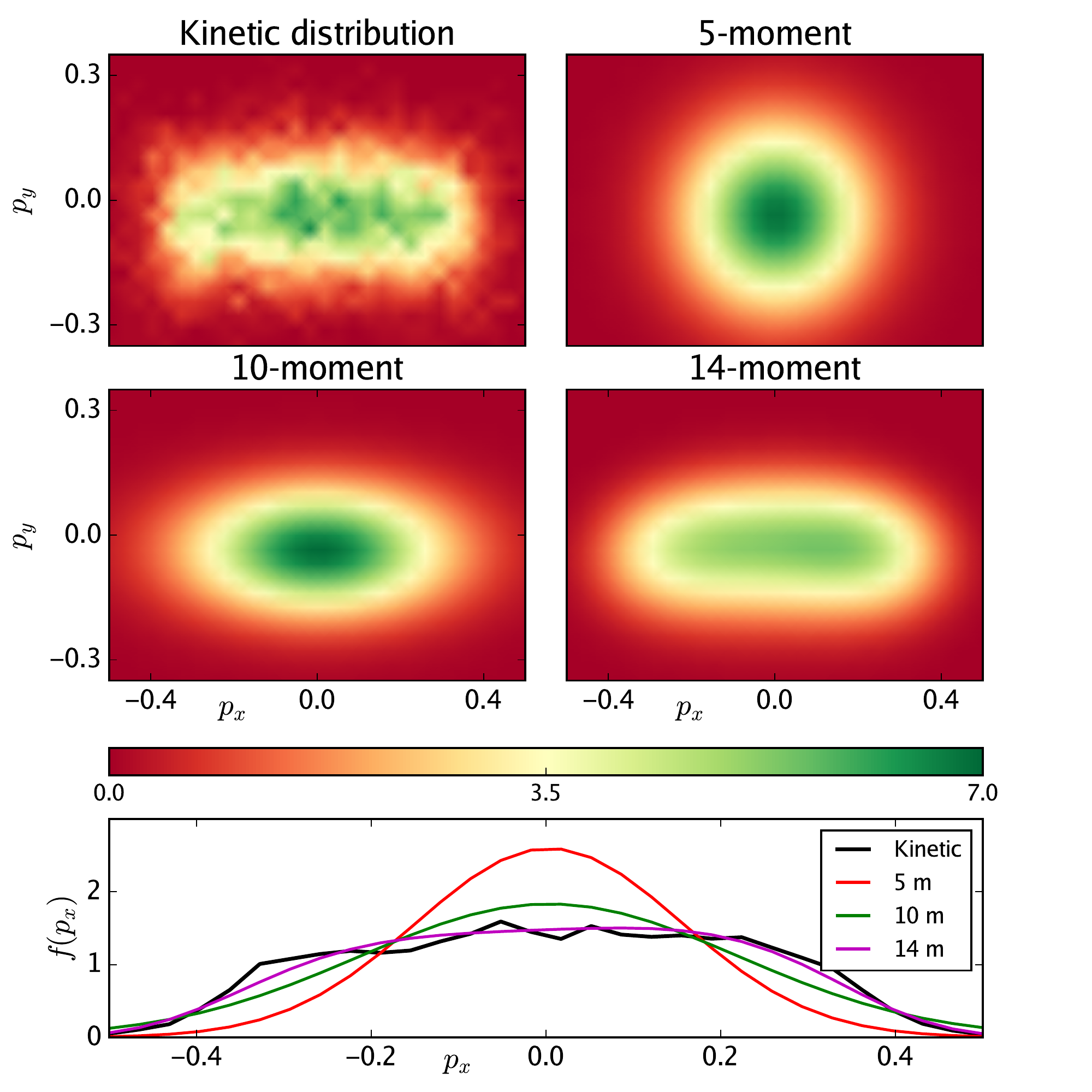}
\caption{Reduced distributions $f(p_x,p_z)$ in the upstream region. Normalisation is such that $\int f(\vec{p}) d^3p = 1$. The lower plot shows the 1-d distribution, highlighting the importance of anisotropy and the $v^4$ moment in capturing the structure of the distribution. The 21-moment distribution is not shown as it is very similar to the 14-moment distribution. }
\label{fig:upstream}
\end{figure}

Just upstream of the electron diffusion region, the electron distributions are anisotropic due to a combination of electrostatic and magnetic trapping \cite{le:2009}. This gives $P_\parallel > P_\perp$, which can be seen in Fig.~\ref{fig:upstream}, where $v_x$ is approximately parallel to the magnetic field. The models with ten or more moments capture the anisotropy correctly, though the 14-moment model better reproduces the small asymmetry and the flat region associated with the trapping, as shown in the one-dimensional plots of the reduced distribution $f(p_x)$. For distributions in this region, using the 21-moment model does not significantly change the reconstruction. 

\begin{figure}
\ig{3.375in}{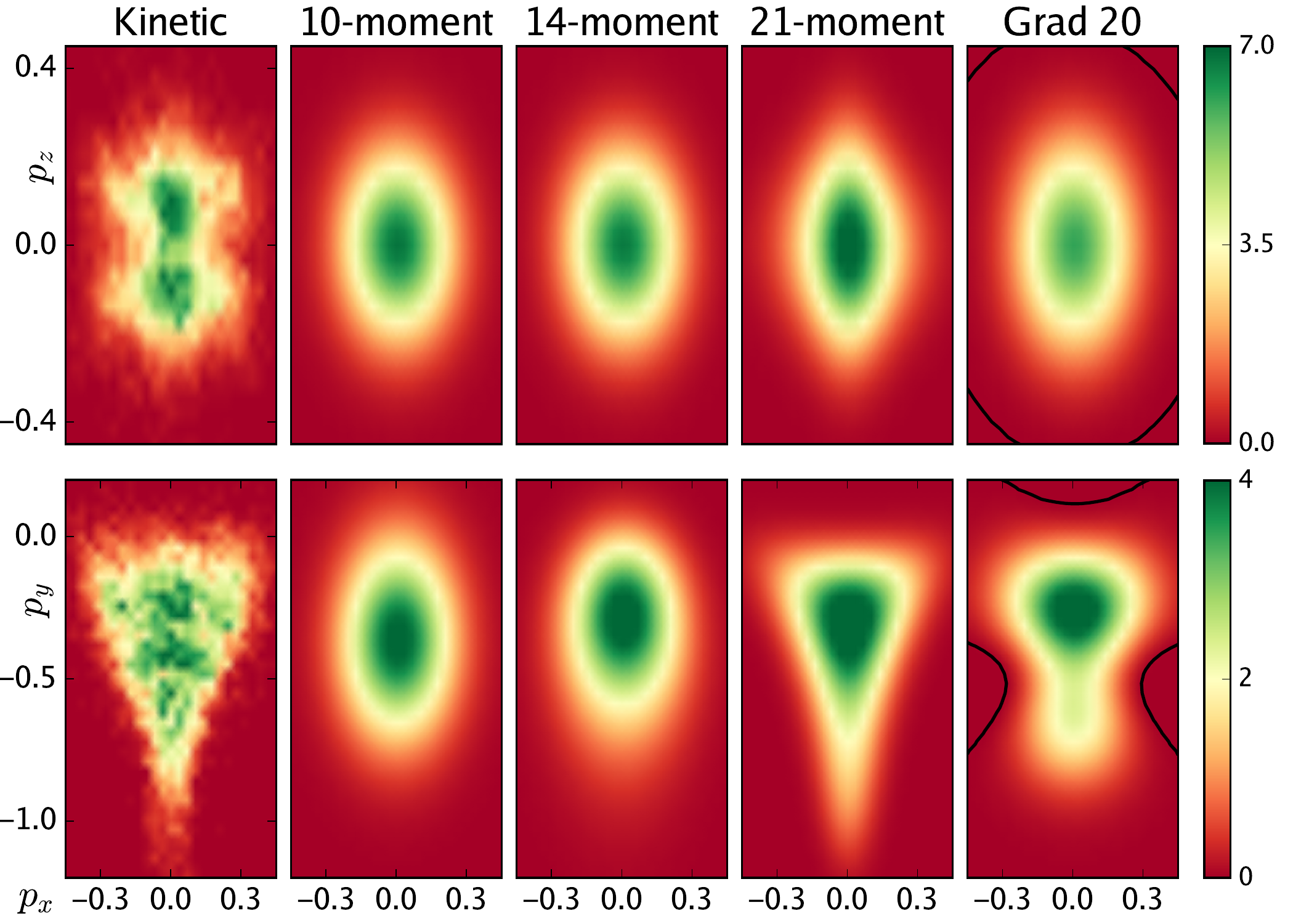}
\caption{Comparison of reduced distribution functions from various models and a kinetic simulation at the x-point. The contours in the rightmost plot show the boundary between regions of positive and negative (unphysical) $f$.}
\label{fig:xpoint}
\end{figure}

The particle distribution at the x-point is shown in the leftmost column Fig.~\ref{fig:xpoint}. The first row shows the reduced distribution $f(p_x,p_z)$, while the second row shows $f(p_x,p_y)$. In the $p_x$-$p_y$ plane, the characteristic triangular shape associated with electron meandering and acceleration in the $p_y$ direction can be seen. The reconstructed distributions are shown in the remaining columns, where the 10- and 14-moment models are unable to reproduce the triangular shape. The 21-moment model is able to capture this shape, but does not have enough detail to capture the bimodal structure in $p_z$ or the finer structures associated with different electron crossings of the current sheet \cite{ng:2011, bessho:2014}. For comparison, a Grad 20-moment distribution is shown in the final column of Fig.~\ref{fig:xpoint}. The elongation in the $p_y$ direction can still be seen, but there are also unphysical regions with negative $f$, whose boundaries are marked by the black contours. 

\begin{figure}
\ig{3.375in}{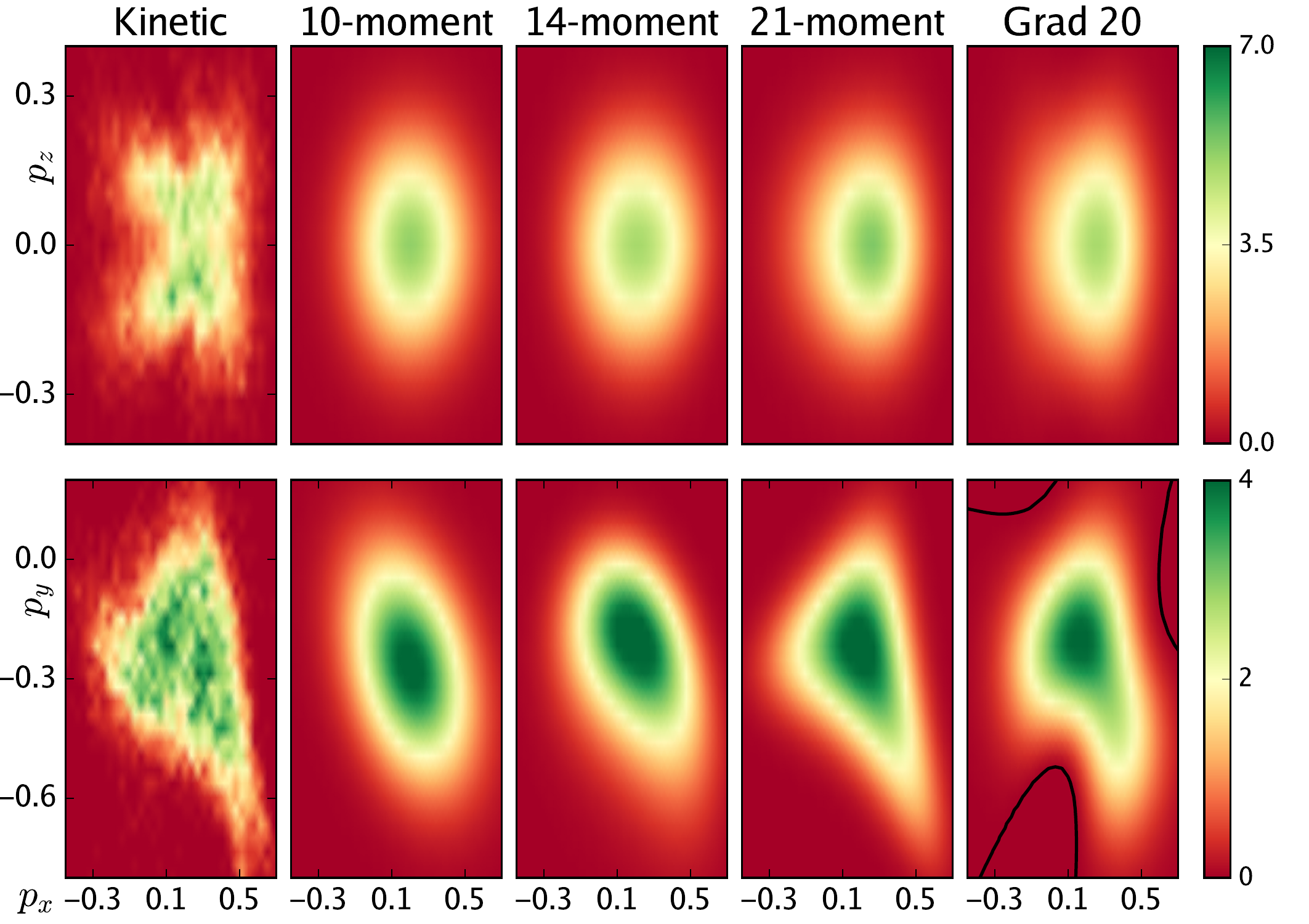}
\caption{Comparison of reduced distribution functions from various models and a kinetic simulation at the 1.7 $d_i$ downstream of the x-point. The contours in the rightmost plot show the boundary between regions of positive and negative (unphysical) $f$.}
\label{fig:downstream}
\end{figure}

Finally, the electron distribution distribution in the exhaust approximately $1.7$ $d_i$ downstream from the x-point is shown in Fig.~\ref{fig:downstream}. Due to the turning of $p_y$ into $p_x$ by $B_z$, the distribution retains its triangular shape but is rotated in the $p_x$-$p_y$ plane compared to the x-point distribution. Again, the effects of adding the various moments can be seen, with the 21-moment model doing the best job of representing the general structure of the kinetic distribution. 

To determine how effective the reconstructions are, we calculate the difference between the model distributions and simulation results using the Hellinger distance \cite{gibbs:2002}
\begin{equation}
H = 1 - \int \sqrt{ f_{Model}(\vec{v}) f(\vec{v})} d^3v.
\label{eq:err}
\end{equation}
In spite of the number of particles being used, there are still some issues with noise affecting the error calculation. Vlasov-Maxwell codes may be a better tool for such comparisons of distributions. For the 20-moment model, where $f$ can become negative in certain regions, we use $|f_{Model}|$ in Eq.~\eqref{eq:err}. 

\begin{figure}
\ig{3.375in}{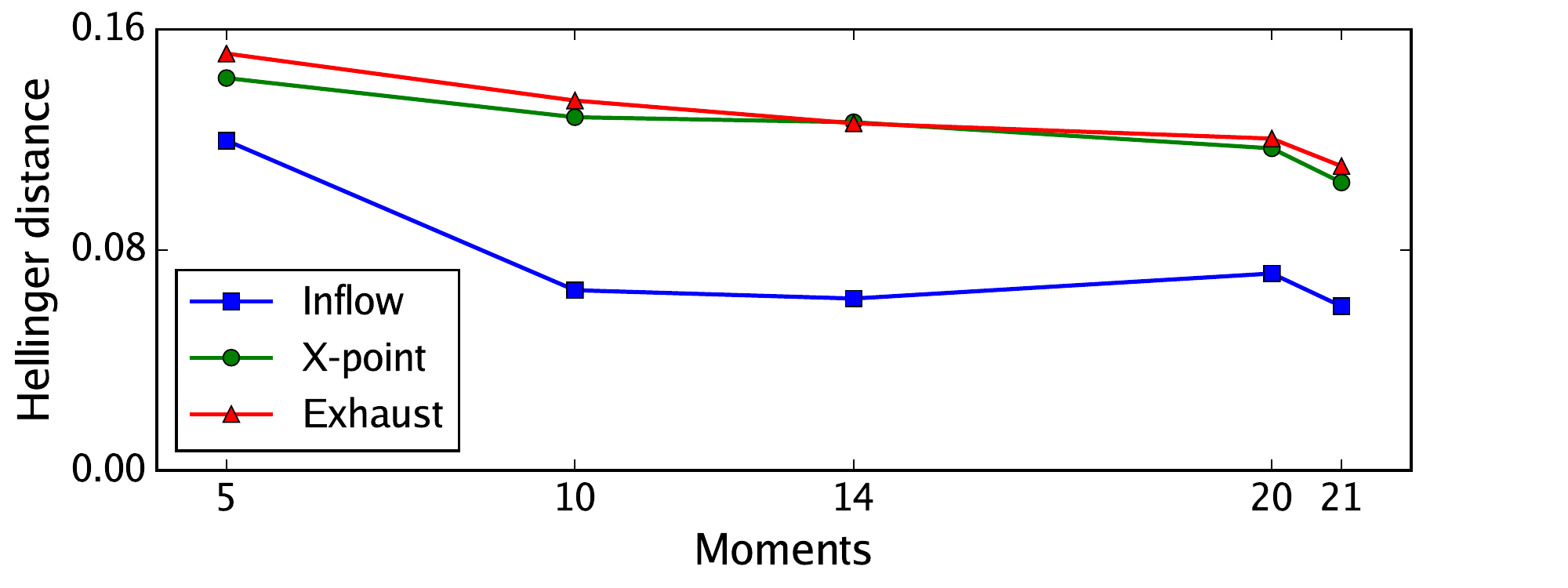}
\caption{Comparison between maximum entropy distributions and particle distributions. }
\label{fig:err}
\end{figure}

The results are shown in Fig.~\ref{fig:err} for the various models and locations in the reconnection region. The agreement between the models increases with the number of moments, though the Grad 20-moment model, with its different functional form, does not necessarily improve the agreement. In the inflow region, the improvement when going from 5 to 10 moments shows the importance of the pressure anisotropy in describing the distributions there, while the improvement going from 10 to 14 and 21 moments is smaller. Due to the location of the sample on the vertical axis of symmetry, the heat flux is small and the deviation from the Gaussian is limited to the flattening of the distribution, which is well described by the scalar $v^4$ moment. At the x-point and in the exhaust, the difference between the model and particle distributions is larger than the inflow case, which is unsurprising due to the larger deviation from the equilibrium distribution and indicates that more moments are needed to fully describe the structure. 

\section{Conclusion} 
\label{sec:conclusion}

We have shown how the maximum entropy method can be used to reconstruct distribution functions from their known moments in a non-perturbative way. Compared to the Grad method, the reconstructed distributions are always positive and guarantee the hyperbolicity of the resulting moment equations, which is important for numerical evaluation. However, a major drawback of this method is the necessity of evaluating the closure coefficients for $>10$ moment models. From a practical standpoint, this computation, which includes velocity space integrals and solving a minimisation problem, must be avoided for affordable fluid simulations to be performed. A possible approach, which has been used for 14-moment neutral fluids \cite{mcdonald:2013interp}, is the development of closed form approximations based on interpolating between the realisability boundaries of the maximum entropy closure, but further development will be necessary for higher moment equations.

This work is focused on understanding the number of moments necessary to capture the structure of the electron distribution function in antiparallel reconnection without additional information. In the inflow region, the Gaussian description of pressure anisotropy is insufficient to describe the flattening of the distribution due to particle trapping, and it is necessary to take into account the $v^4$ moment. 

Within the current sheet, the more complicated distribution functions are not well described by even the 21-moment model, which only captures the electron acceleration in the $y$ direction without the structure associated with meandering and counter-streaming particles. The description of the counter-streaming flows likely requires retaining more of the fourth order moments (i.e.~the 26- or 35-moment models) \cite{schaerer:2017}, while the finer scale structures with multiple electron populations possibly require many more moments, based on the results of a similar closure model in 1-D, where 15 moments were required to properly describe a distribution with three peaks \cite{abdelmalik:2016}. 

Because the maximum entropy model closure is based only on the known fluid moments, it does not  require or provide information about the physical processes involved in the evolution of the distribution function. Using the reconnection inflow as an example again, the anisotropy due to trapping can be described by equations of state for $P_\parallel$ and $P_\perp$ \cite{le:2009}, while the maximum entropy closure requires 14 moments. On the other hand, the equations of state require knowledge of the background plasma density and magnetic field, while the maximum entropy closure requires only local information. Thus, while it is less efficient compared to specific physical models, the same form of the distribution and set of equations can be used more generally. 

Finally, we comment briefly on the relation of these results to existing ten-moment descriptions of reconnection. In spite of the differences between the distribution functions, ten-moment methods with an approximate heat flux or temperature relaxation have ostensibly had some success in modeling reconnection \cite{hesse:1995,yin:2001,wang:2015,ng:2017}. The balancing of the reconnection electric field by the pressure tensor is included, as are some aspects of the anisotropy, but the importance of the heat flux (which is not fully evaluated by some of these models) is seen in the sensitivity of the reconnection rates to the closure approximations \cite{wang:2015,ng:2015}. 

This work was supported by DOE Contract DE-AC02-09CH11466 and NSF Grant AGS-1338944.

\bibliography{reconnectionbib}

\begin{thebibliography}{46}%
\makeatletter
\providecommand \@ifxundefined [1]{%
 \@ifx{#1\undefined}
}%
\providecommand \@ifnum [1]{%
 \ifnum #1\expandafter \@firstoftwo
 \else \expandafter \@secondoftwo
 \fi
}%
\providecommand \@ifx [1]{%
 \ifx #1\expandafter \@firstoftwo
 \else \expandafter \@secondoftwo
 \fi
}%
\providecommand \natexlab [1]{#1}%
\providecommand \enquote  [1]{``#1''}%
\providecommand \bibnamefont  [1]{#1}%
\providecommand \bibfnamefont [1]{#1}%
\providecommand \citenamefont [1]{#1}%
\providecommand \href@noop [0]{\@secondoftwo}%
\providecommand \href [0]{\begingroup \@sanitize@url \@href}%
\providecommand \@href[1]{\@@startlink{#1}\@@href}%
\providecommand \@@href[1]{\endgroup#1\@@endlink}%
\providecommand \@sanitize@url [0]{\catcode `\\12\catcode `\$12\catcode
  `\&12\catcode `\#12\catcode `\^12\catcode `\_12\catcode `\%12\relax}%
\providecommand \@@startlink[1]{}%
\providecommand \@@endlink[0]{}%
\providecommand \url  [0]{\begingroup\@sanitize@url \@url }%
\providecommand \@url [1]{\endgroup\@href {#1}{\urlprefix }}%
\providecommand \urlprefix  [0]{URL }%
\providecommand \Eprint [0]{\href }%
\providecommand \doibase [0]{http://dx.doi.org/}%
\providecommand \selectlanguage [0]{\@gobble}%
\providecommand \bibinfo  [0]{\@secondoftwo}%
\providecommand \bibfield  [0]{\@secondoftwo}%
\providecommand \translation [1]{[#1]}%
\providecommand \BibitemOpen [0]{}%
\providecommand \bibitemStop [0]{}%
\providecommand \bibitemNoStop [0]{.\EOS\space}%
\providecommand \EOS [0]{\spacefactor3000\relax}%
\providecommand \BibitemShut  [1]{\csname bibitem#1\endcsname}%
\let\auto@bib@innerbib\@empty
\bibitem [{\citenamefont {Marsch}(2006)}]{marsch:2006}%
  \BibitemOpen
  \bibfield  {author} {\bibinfo {author} {\bibfnamefont {E.}~\bibnamefont
  {Marsch}},\ }\href {\doibase 10.12942/lrsp-2006-1} {\bibfield  {journal}
  {\bibinfo  {journal} {Living Reviews in Solar Physics}\ }\textbf {\bibinfo
  {volume} {3}},\ \bibinfo {pages} {1} (\bibinfo {year} {2006})}\BibitemShut
  {NoStop}%
\bibitem [{\citenamefont {Burch}\ and\ \citenamefont
  {Phan}(2016)}]{burch:2016}%
  \BibitemOpen
  \bibfield  {author} {\bibinfo {author} {\bibfnamefont {J.~L.}\ \bibnamefont
  {Burch}}\ and\ \bibinfo {author} {\bibfnamefont {T.~D.}\ \bibnamefont
  {Phan}},\ }\href {\doibase 10.1002/2016GL069787} {\bibfield  {journal}
  {\bibinfo  {journal} {Geophysical Research Letters}\ }\textbf {\bibinfo
  {volume} {43}},\ \bibinfo {pages} {8327} (\bibinfo {year} {2016})},\ \bibinfo
  {note} {2016GL069787}\BibitemShut {NoStop}%
\bibitem [{\citenamefont {Oka}\ \emph {et~al.}(2016)\citenamefont {Oka},
  \citenamefont {Phan}, \citenamefont {Øieroset},\ and\ \citenamefont
  {Angelopoulos}}]{oka:2016}%
  \BibitemOpen
  \bibfield  {author} {\bibinfo {author} {\bibfnamefont {M.}~\bibnamefont
  {Oka}}, \bibinfo {author} {\bibfnamefont {T.-D.}\ \bibnamefont {Phan}},
  \bibinfo {author} {\bibfnamefont {M.}~\bibnamefont {Øieroset}}, \ and\
  \bibinfo {author} {\bibfnamefont {V.}~\bibnamefont {Angelopoulos}},\ }\href
  {\doibase 10.1002/2015JA022040} {\bibfield  {journal} {\bibinfo  {journal}
  {Journal of Geophysical Research: Space Physics}\ }\textbf {\bibinfo {volume}
  {121}},\ \bibinfo {pages} {1955} (\bibinfo {year} {2016})},\ \bibinfo {note}
  {2015JA022040}\BibitemShut {NoStop}%
\bibitem [{\citenamefont {Dungey}(1953)}]{dungey:1953}%
  \BibitemOpen
  \bibfield  {author} {\bibinfo {author} {\bibfnamefont {J.}~\bibnamefont
  {Dungey}},\ }\href {\doibase 10.1080/14786440708521050} {\bibfield  {journal}
  {\bibinfo  {journal} {Philosophical Magazine Series 7}\ }\textbf {\bibinfo
  {volume} {44}},\ \bibinfo {pages} {725} (\bibinfo {year} {1953})}\BibitemShut
  {NoStop}%
\bibitem [{\citenamefont {Le}\ \emph {et~al.}(2009)\citenamefont {Le},
  \citenamefont {Egedal}, \citenamefont {Daughton}, \citenamefont {Fox},\ and\
  \citenamefont {Katz}}]{le:2009}%
  \BibitemOpen
  \bibfield  {author} {\bibinfo {author} {\bibfnamefont {A.}~\bibnamefont
  {Le}}, \bibinfo {author} {\bibfnamefont {J.}~\bibnamefont {Egedal}}, \bibinfo
  {author} {\bibfnamefont {W.}~\bibnamefont {Daughton}}, \bibinfo {author}
  {\bibfnamefont {W.}~\bibnamefont {Fox}}, \ and\ \bibinfo {author}
  {\bibfnamefont {N.}~\bibnamefont {Katz}},\ }\href {\doibase
  {10.1103/PhysRevLett.102.085001}} {\bibfield  {journal} {\bibinfo  {journal}
  {Phys. Rev. Lett}\ }\textbf {\bibinfo {volume} {{102}}},\ \bibinfo {pages}
  {{085001}} (\bibinfo {year} {2009})}\BibitemShut {NoStop}%
\bibitem [{\citenamefont {Bessho}\ \emph {et~al.}(2014)\citenamefont {Bessho},
  \citenamefont {Chen}, \citenamefont {Shuster},\ and\ \citenamefont
  {Wang}}]{bessho:2014}%
  \BibitemOpen
  \bibfield  {author} {\bibinfo {author} {\bibfnamefont {N.}~\bibnamefont
  {Bessho}}, \bibinfo {author} {\bibfnamefont {L.-J.}\ \bibnamefont {Chen}},
  \bibinfo {author} {\bibfnamefont {J.~R.}\ \bibnamefont {Shuster}}, \ and\
  \bibinfo {author} {\bibfnamefont {S.}~\bibnamefont {Wang}},\ }\href {\doibase
  10.1002/2014GL062034} {\bibfield  {journal} {\bibinfo  {journal} {Geophysical
  Research Letters}\ }\textbf {\bibinfo {volume} {41}},\ \bibinfo {pages}
  {8688} (\bibinfo {year} {2014})}\BibitemShut {NoStop}%
\bibitem [{\citenamefont {Bessho}\ \emph {et~al.}(2017)\citenamefont {Bessho},
  \citenamefont {Chen}, \citenamefont {Hesse},\ and\ \citenamefont
  {Wang}}]{bessho:2017}%
  \BibitemOpen
  \bibfield  {author} {\bibinfo {author} {\bibfnamefont {N.}~\bibnamefont
  {Bessho}}, \bibinfo {author} {\bibfnamefont {L.-J.}\ \bibnamefont {Chen}},
  \bibinfo {author} {\bibfnamefont {M.}~\bibnamefont {Hesse}}, \ and\ \bibinfo
  {author} {\bibfnamefont {S.}~\bibnamefont {Wang}},\ }\href {\doibase
  10.1063/1.4989737} {\bibfield  {journal} {\bibinfo  {journal} {Physics of
  Plasmas}\ }\textbf {\bibinfo {volume} {24}},\ \bibinfo {pages} {072903}
  (\bibinfo {year} {2017})},\ \Eprint
  {http://arxiv.org/abs/https://doi.org/10.1063/1.4989737}
  {https://doi.org/10.1063/1.4989737} \BibitemShut {NoStop}%
\bibitem [{\citenamefont {Yin}\ \emph {et~al.}(2001)\citenamefont {Yin},
  \citenamefont {Winske}, \citenamefont {Gary},\ and\ \citenamefont
  {Birn}}]{yin:2001}%
  \BibitemOpen
  \bibfield  {author} {\bibinfo {author} {\bibfnamefont {L.}~\bibnamefont
  {Yin}}, \bibinfo {author} {\bibfnamefont {D.}~\bibnamefont {Winske}},
  \bibinfo {author} {\bibfnamefont {S.~P.}\ \bibnamefont {Gary}}, \ and\
  \bibinfo {author} {\bibfnamefont {J.}~\bibnamefont {Birn}},\ }\href {\doibase
  10.1029/2000JA000398} {\bibfield  {journal} {\bibinfo  {journal} {Journal of
  Geophysical Research: Space Physics}\ }\textbf {\bibinfo {volume} {106}},\
  \bibinfo {pages} {10761} (\bibinfo {year} {2001})}\BibitemShut {NoStop}%
\bibitem [{\citenamefont {Wang}\ \emph {et~al.}(2015)\citenamefont {Wang},
  \citenamefont {Hakim}, \citenamefont {Bhattacharjee},\ and\ \citenamefont
  {Germaschewski}}]{wang:2015}%
  \BibitemOpen
  \bibfield  {author} {\bibinfo {author} {\bibfnamefont {L.}~\bibnamefont
  {Wang}}, \bibinfo {author} {\bibfnamefont {A.~H.}\ \bibnamefont {Hakim}},
  \bibinfo {author} {\bibfnamefont {A.}~\bibnamefont {Bhattacharjee}}, \ and\
  \bibinfo {author} {\bibfnamefont {K.}~\bibnamefont {Germaschewski}},\ }\href
  {\doibase http://dx.doi.org/10.1063/1.4906063} {\bibfield  {journal}
  {\bibinfo  {journal} {Physics of Plasmas (1994-present)}\ }\textbf {\bibinfo
  {volume} {22}},\ \bibinfo {eid} {012108} (\bibinfo {year}
  {2015})}\BibitemShut {NoStop}%
\bibitem [{\citenamefont {Miller}\ and\ \citenamefont
  {Shumlak}(2016)}]{miller:2016}%
  \BibitemOpen
  \bibfield  {author} {\bibinfo {author} {\bibfnamefont {S.~T.}\ \bibnamefont
  {Miller}}\ and\ \bibinfo {author} {\bibfnamefont {U.}~\bibnamefont
  {Shumlak}},\ }\href {\doibase 10.1063/1.4960041} {\bibfield  {journal}
  {\bibinfo  {journal} {Physics of Plasmas}\ }\textbf {\bibinfo {volume}
  {23}},\ \bibinfo {pages} {082303} (\bibinfo {year} {2016})},\ \Eprint
  {http://arxiv.org/abs/http://dx.doi.org/10.1063/1.4960041}
  {http://dx.doi.org/10.1063/1.4960041} \BibitemShut {NoStop}%
\bibitem [{\citenamefont {Ng}\ \emph {et~al.}(2017)\citenamefont {Ng},
  \citenamefont {Hakim}, \citenamefont {Bhattacharjee}, \citenamefont
  {Stanier},\ and\ \citenamefont {Daughton}}]{ng:2017}%
  \BibitemOpen
  \bibfield  {author} {\bibinfo {author} {\bibfnamefont {J.}~\bibnamefont
  {Ng}}, \bibinfo {author} {\bibfnamefont {A.}~\bibnamefont {Hakim}}, \bibinfo
  {author} {\bibfnamefont {A.}~\bibnamefont {Bhattacharjee}}, \bibinfo {author}
  {\bibfnamefont {A.}~\bibnamefont {Stanier}}, \ and\ \bibinfo {author}
  {\bibfnamefont {W.}~\bibnamefont {Daughton}},\ }\href {\doibase
  10.1063/1.4993195} {\bibfield  {journal} {\bibinfo  {journal} {Physics of
  Plasmas}\ }\textbf {\bibinfo {volume} {24}},\ \bibinfo {pages} {082112}
  (\bibinfo {year} {2017})},\ \Eprint
  {http://arxiv.org/abs/https://doi.org/10.1063/1.4993195}
  {https://doi.org/10.1063/1.4993195} \BibitemShut {NoStop}%
\bibitem [{\citenamefont {Jaynes}(1957)}]{jaynes:1957}%
  \BibitemOpen
  \bibfield  {author} {\bibinfo {author} {\bibfnamefont {E.~T.}\ \bibnamefont
  {Jaynes}},\ }\href {\doibase 10.1103/PhysRev.106.620} {\bibfield  {journal}
  {\bibinfo  {journal} {Phys. Rev.}\ }\textbf {\bibinfo {volume} {106}},\
  \bibinfo {pages} {620} (\bibinfo {year} {1957})}\BibitemShut {NoStop}%
\bibitem [{\citenamefont {{Richstone}}\ and\ \citenamefont
  {{Tremaine}}(1988)}]{richstone:1988}%
  \BibitemOpen
  \bibfield  {author} {\bibinfo {author} {\bibfnamefont {D.~O.}\ \bibnamefont
  {{Richstone}}}\ and\ \bibinfo {author} {\bibfnamefont {S.}~\bibnamefont
  {{Tremaine}}},\ }\href {\doibase 10.1086/166171} {\bibfield  {journal}
  {\bibinfo  {journal} {The Astrophysical Journal}\ }\textbf {\bibinfo {volume}
  {327}},\ \bibinfo {pages} {82} (\bibinfo {year} {1988})}\BibitemShut
  {NoStop}%
\bibitem [{\citenamefont {Yeo}\ and\ \citenamefont {Burge}(2004)}]{yeo:2004}%
  \BibitemOpen
  \bibfield  {author} {\bibinfo {author} {\bibfnamefont {G.}~\bibnamefont
  {Yeo}}\ and\ \bibinfo {author} {\bibfnamefont {C.~B.}\ \bibnamefont
  {Burge}},\ }\href {\doibase 10.1089/1066527041410418} {\bibfield  {journal}
  {\bibinfo  {journal} {Journal of Computational Biology}\ }\textbf {\bibinfo
  {volume} {11}},\ \bibinfo {pages} {377} (\bibinfo {year} {2004})},\ \Eprint
  {http://arxiv.org/abs/https://doi.org/10.1089/1066527041410418}
  {https://doi.org/10.1089/1066527041410418} \BibitemShut {NoStop}%
\bibitem [{\citenamefont {Charniak}(2000)}]{charniak:2000}%
  \BibitemOpen
  \bibfield  {author} {\bibinfo {author} {\bibfnamefont {E.}~\bibnamefont
  {Charniak}},\ }in\ \href {http://dl.acm.org/citation.cfm?id=974305.974323}
  {\emph {\bibinfo {booktitle} {Proceedings of the 1st North American Chapter
  of the Association for Computational Linguistics Conference}}},\ \bibinfo
  {series and number} {NAACL 2000}\ (\bibinfo  {publisher} {Association for
  Computational Linguistics},\ \bibinfo {address} {Stroudsburg, PA, USA},\
  \bibinfo {year} {2000})\ pp.\ \bibinfo {pages} {132--139}\BibitemShut
  {NoStop}%
\bibitem [{\citenamefont {Levermore}(1996)}]{levermore:1996}%
  \BibitemOpen
  \bibfield  {author} {\bibinfo {author} {\bibfnamefont {C.~D.}\ \bibnamefont
  {Levermore}},\ }\href {\doibase 10.1007/BF02179552} {\bibfield  {journal}
  {\bibinfo  {journal} {Journal of Statistical Physics}\ }\textbf {\bibinfo
  {volume} {83}},\ \bibinfo {pages} {1021} (\bibinfo {year}
  {1996})}\BibitemShut {NoStop}%
\bibitem [{\citenamefont {Grad}(1949{\natexlab{a}})}]{grad:1949}%
  \BibitemOpen
  \bibfield  {author} {\bibinfo {author} {\bibfnamefont {H.}~\bibnamefont
  {Grad}},\ }\href {\doibase 10.1002/cpa.3160020403} {\bibfield  {journal}
  {\bibinfo  {journal} {Communications on Pure and Applied Mathematics}\
  }\textbf {\bibinfo {volume} {2}},\ \bibinfo {pages} {331} (\bibinfo {year}
  {1949}{\natexlab{a}})}\BibitemShut {NoStop}%
\bibitem [{\citenamefont {Chapman}\ and\ \citenamefont
  {Cowling}(1970)}]{chapman:1970}%
  \BibitemOpen
  \bibfield  {author} {\bibinfo {author} {\bibfnamefont {S.}~\bibnamefont
  {Chapman}}\ and\ \bibinfo {author} {\bibfnamefont {T.~G.}\ \bibnamefont
  {Cowling}},\ }\href@noop {} {\emph {\bibinfo {title} {The mathematical theory
  of non-uniform gases: an account of the kinetic theory of viscosity, thermal
  conduction and diffusion in gases}}}\ (\bibinfo  {publisher} {Cambridge
  university press},\ \bibinfo {year} {1970})\BibitemShut {NoStop}%
\bibitem [{\citenamefont {Bobylev}(2008)}]{bobylev:2008}%
  \BibitemOpen
  \bibfield  {author} {\bibinfo {author} {\bibfnamefont {A.~V.}\ \bibnamefont
  {Bobylev}},\ }\href {\doibase 10.1007/s10955-008-9556-5} {\bibfield
  {journal} {\bibinfo  {journal} {Journal of Statistical Physics}\ }\textbf
  {\bibinfo {volume} {132}},\ \bibinfo {pages} {569} (\bibinfo {year}
  {2008})}\BibitemShut {NoStop}%
\bibitem [{\citenamefont {Grad}(1949{\natexlab{b}})}]{grad:1949hermite}%
  \BibitemOpen
  \bibfield  {author} {\bibinfo {author} {\bibfnamefont {H.}~\bibnamefont
  {Grad}},\ }\href {\doibase 10.1002/cpa.3160020402} {\bibfield  {journal}
  {\bibinfo  {journal} {Communications on Pure and Applied Mathematics}\
  }\textbf {\bibinfo {volume} {2}},\ \bibinfo {pages} {325} (\bibinfo {year}
  {1949}{\natexlab{b}})}\BibitemShut {NoStop}%
\bibitem [{\citenamefont {Cai}, \citenamefont {Fan},\ and\ \citenamefont
  {Li}(2015)}]{cai:2015}%
  \BibitemOpen
  \bibfield  {author} {\bibinfo {author} {\bibfnamefont {Z.}~\bibnamefont
  {Cai}}, \bibinfo {author} {\bibfnamefont {Y.}~\bibnamefont {Fan}}, \ and\
  \bibinfo {author} {\bibfnamefont {R.}~\bibnamefont {Li}},\ }\href {\doibase
  10.1137/14100110X} {\bibfield  {journal} {\bibinfo  {journal} {SIAM Journal
  on Applied Mathematics}\ }\textbf {\bibinfo {volume} {75}},\ \bibinfo {pages}
  {2001} (\bibinfo {year} {2015})},\ \Eprint
  {http://arxiv.org/abs/http://dx.doi.org/10.1137/14100110X}
  {http://dx.doi.org/10.1137/14100110X} \BibitemShut {NoStop}%
\bibitem [{\citenamefont {Struchtrup}\ and\ \citenamefont
  {Torrilhon}(2003)}]{struchtrup:2003}%
  \BibitemOpen
  \bibfield  {author} {\bibinfo {author} {\bibfnamefont {H.}~\bibnamefont
  {Struchtrup}}\ and\ \bibinfo {author} {\bibfnamefont {M.}~\bibnamefont
  {Torrilhon}},\ }\href {\doibase http://dx.doi.org/10.1063/1.1597472}
  {\bibfield  {journal} {\bibinfo  {journal} {Physics of Fluids}\ }\textbf
  {\bibinfo {volume} {15}},\ \bibinfo {pages} {2668} (\bibinfo {year}
  {2003})}\BibitemShut {NoStop}%
\bibitem [{\citenamefont {Torrilhon}\ and\ \citenamefont
  {Struchtrup}(2004)}]{torrilhon:2004}%
  \BibitemOpen
  \bibfield  {author} {\bibinfo {author} {\bibfnamefont {M.}~\bibnamefont
  {Torrilhon}}\ and\ \bibinfo {author} {\bibfnamefont {H.}~\bibnamefont
  {Struchtrup}},\ }\href {\doibase 10.1017/S0022112004009917} {\bibfield
  {journal} {\bibinfo  {journal} {Journal of Fluid Mechanics}\ }\textbf
  {\bibinfo {volume} {513}},\ \bibinfo {pages} {171–198} (\bibinfo {year}
  {2004})}\BibitemShut {NoStop}%
\bibitem [{\citenamefont {Torrilhon}(2010)}]{torrilhon:2010}%
  \BibitemOpen
  \bibfield  {author} {\bibinfo {author} {\bibfnamefont {M.}~\bibnamefont
  {Torrilhon}},\ }\href@noop {} {\bibfield  {journal} {\bibinfo  {journal}
  {Commun. Comput. Phys}\ }\textbf {\bibinfo {volume} {7}},\ \bibinfo {pages}
  {639} (\bibinfo {year} {2010})}\BibitemShut {NoStop}%
\bibitem [{\citenamefont {\"Ottinger}(2010)}]{ottinger:2010}%
  \BibitemOpen
  \bibfield  {author} {\bibinfo {author} {\bibfnamefont {H.~C.}\ \bibnamefont
  {\"Ottinger}},\ }\href {\doibase 10.1103/PhysRevLett.104.120601} {\bibfield
  {journal} {\bibinfo  {journal} {Phys. Rev. Lett.}\ }\textbf {\bibinfo
  {volume} {104}},\ \bibinfo {pages} {120601} (\bibinfo {year}
  {2010})}\BibitemShut {NoStop}%
\bibitem [{\citenamefont {Singh}\ and\ \citenamefont
  {Agrawal}(2016)}]{singh:2016}%
  \BibitemOpen
  \bibfield  {author} {\bibinfo {author} {\bibfnamefont {N.}~\bibnamefont
  {Singh}}\ and\ \bibinfo {author} {\bibfnamefont {A.}~\bibnamefont
  {Agrawal}},\ }\href {\doibase 10.1103/PhysRevE.93.063111} {\bibfield
  {journal} {\bibinfo  {journal} {Phys. Rev. E}\ }\textbf {\bibinfo {volume}
  {93}},\ \bibinfo {pages} {063111} (\bibinfo {year} {2016})}\BibitemShut
  {NoStop}%
\bibitem [{\citenamefont {{Junk}}(1998)}]{junk:1998}%
  \BibitemOpen
  \bibfield  {author} {\bibinfo {author} {\bibfnamefont {M.}~\bibnamefont
  {{Junk}}},\ }\href {\doibase 10.1023/B:JOSS.0000033155.07331.d9} {\bibfield
  {journal} {\bibinfo  {journal} {Journal of Statistical Physics}\ }\textbf
  {\bibinfo {volume} {93}},\ \bibinfo {pages} {1143} (\bibinfo {year}
  {1998})}\BibitemShut {NoStop}%
\bibitem [{\citenamefont {McDonald}\ and\ \citenamefont
  {Groth}(2013)}]{mcdonald:2013}%
  \BibitemOpen
  \bibfield  {author} {\bibinfo {author} {\bibfnamefont {J.~G.}\ \bibnamefont
  {McDonald}}\ and\ \bibinfo {author} {\bibfnamefont {C.~P.~T.}\ \bibnamefont
  {Groth}},\ }\href {\doibase 10.1007/s00161-012-0252-y} {\bibfield  {journal}
  {\bibinfo  {journal} {Continuum Mechanics and Thermodynamics}\ }\textbf
  {\bibinfo {volume} {25}},\ \bibinfo {pages} {573} (\bibinfo {year}
  {2013})}\BibitemShut {NoStop}%
\bibitem [{\citenamefont {Groth}\ and\ \citenamefont
  {McDonald}(2009)}]{groth:2009}%
  \BibitemOpen
  \bibfield  {author} {\bibinfo {author} {\bibfnamefont {C.~P.~T.}\
  \bibnamefont {Groth}}\ and\ \bibinfo {author} {\bibfnamefont {J.~G.}\
  \bibnamefont {McDonald}},\ }\href {\doibase 10.1007/s00161-009-0125-1}
  {\bibfield  {journal} {\bibinfo  {journal} {Continuum Mechanics and
  Thermodynamics}\ }\textbf {\bibinfo {volume} {21}},\ \bibinfo {pages} {467}
  (\bibinfo {year} {2009})}\BibitemShut {NoStop}%
\bibitem [{\citenamefont {McDonald}\ and\ \citenamefont
  {Torrilhon}(2013)}]{mcdonald:2013interp}%
  \BibitemOpen
  \bibfield  {author} {\bibinfo {author} {\bibfnamefont {J.}~\bibnamefont
  {McDonald}}\ and\ \bibinfo {author} {\bibfnamefont {M.}~\bibnamefont
  {Torrilhon}},\ }\href {\doibase https://doi.org/10.1016/j.jcp.2013.05.046}
  {\bibfield  {journal} {\bibinfo  {journal} {Journal of Computational
  Physics}\ }\textbf {\bibinfo {volume} {251}},\ \bibinfo {pages} {500 }
  (\bibinfo {year} {2013})}\BibitemShut {NoStop}%
\bibitem [{\citenamefont {Tensuda}, \citenamefont {McDonald},\ and\
  \citenamefont {Groth}(2016)}]{boone:2016}%
  \BibitemOpen
  \bibfield  {author} {\bibinfo {author} {\bibfnamefont {B.~R.}\ \bibnamefont
  {Tensuda}}, \bibinfo {author} {\bibfnamefont {J.~G.}\ \bibnamefont
  {McDonald}}, \ and\ \bibinfo {author} {\bibfnamefont {C.~P.~T.}\ \bibnamefont
  {Groth}},\ }\href {\doibase 10.1063/1.4967639} {\bibfield  {journal}
  {\bibinfo  {journal} {AIP Conference Proceedings}\ }\textbf {\bibinfo
  {volume} {1786}},\ \bibinfo {pages} {140008} (\bibinfo {year} {2016})},\
  \Eprint
  {http://arxiv.org/abs/https://aip.scitation.org/doi/pdf/10.1063/1.4967639}
  {https://aip.scitation.org/doi/pdf/10.1063/1.4967639} \BibitemShut {NoStop}%
\bibitem [{\citenamefont {Schaerer}\ and\ \citenamefont
  {Torrilhon}(2017)}]{schaerer:2017}%
  \BibitemOpen
  \bibfield  {author} {\bibinfo {author} {\bibfnamefont {R.~P.}\ \bibnamefont
  {Schaerer}}\ and\ \bibinfo {author} {\bibfnamefont {M.}~\bibnamefont
  {Torrilhon}},\ }\href {\doibase
  https://doi.org/10.1016/j.euromechflu.2017.01.003} {\bibfield  {journal}
  {\bibinfo  {journal} {European Journal of Mechanics - B/Fluids}\ }\textbf
  {\bibinfo {volume} {64}},\ \bibinfo {pages} {30 } (\bibinfo {year} {2017})},\
  \bibinfo {note} {special Issue on Non-equilibrium Gas Flows}\BibitemShut
  {NoStop}%
\bibitem [{\citenamefont {Chen}\ \emph {et~al.}(2008)\citenamefont {Chen},
  \citenamefont {Bessho}, \citenamefont {Lefebvre}, \citenamefont {Vaith},
  \citenamefont {Fazakerley}, \citenamefont {Bhattacharjee}, \citenamefont
  {Puhl-Quinn}, \citenamefont {Runov}, \citenamefont {Khotyaintsev},
  \citenamefont {Vaivads}, \citenamefont {Georgescu},\ and\ \citenamefont
  {Torbert}}]{chen:2008}%
  \BibitemOpen
  \bibfield  {author} {\bibinfo {author} {\bibfnamefont {L.-J.}\ \bibnamefont
  {Chen}}, \bibinfo {author} {\bibfnamefont {N.}~\bibnamefont {Bessho}},
  \bibinfo {author} {\bibfnamefont {B.}~\bibnamefont {Lefebvre}}, \bibinfo
  {author} {\bibfnamefont {H.}~\bibnamefont {Vaith}}, \bibinfo {author}
  {\bibfnamefont {A.}~\bibnamefont {Fazakerley}}, \bibinfo {author}
  {\bibfnamefont {A.}~\bibnamefont {Bhattacharjee}}, \bibinfo {author}
  {\bibfnamefont {P.~A.}\ \bibnamefont {Puhl-Quinn}}, \bibinfo {author}
  {\bibfnamefont {A.}~\bibnamefont {Runov}}, \bibinfo {author} {\bibfnamefont
  {Y.}~\bibnamefont {Khotyaintsev}}, \bibinfo {author} {\bibfnamefont
  {A.}~\bibnamefont {Vaivads}}, \bibinfo {author} {\bibfnamefont
  {E.}~\bibnamefont {Georgescu}}, \ and\ \bibinfo {author} {\bibfnamefont
  {R.}~\bibnamefont {Torbert}},\ }\href {\doibase 10.1029/2008JA013385}
  {\bibfield  {journal} {\bibinfo  {journal} {Journal of Geophysical Research:
  Space Physics}\ }\textbf {\bibinfo {volume} {113}} (\bibinfo {year} {2008}),\
  10.1029/2008JA013385}\BibitemShut {NoStop}%
\bibitem [{\citenamefont {Egedal}, \citenamefont {Le},\ and\ \citenamefont
  {Daughton}(2013)}]{egedal:2013}%
  \BibitemOpen
  \bibfield  {author} {\bibinfo {author} {\bibfnamefont {J.}~\bibnamefont
  {Egedal}}, \bibinfo {author} {\bibfnamefont {A.}~\bibnamefont {Le}}, \ and\
  \bibinfo {author} {\bibfnamefont {W.}~\bibnamefont {Daughton}},\ }\href
  {\doibase http://dx.doi.org/10.1063/1.4811092} {\bibfield  {journal}
  {\bibinfo  {journal} {Physics of Plasmas}\ }\textbf {\bibinfo {volume}
  {20}},\ \bibinfo {eid} {061201} (\bibinfo {year} {2013})}\BibitemShut
  {NoStop}%
\bibitem [{\citenamefont {Le}\ \emph {et~al.}(2017)\citenamefont {Le},
  \citenamefont {Daughton}, \citenamefont {Chen},\ and\ \citenamefont
  {Egedal}}]{le:2017}%
  \BibitemOpen
  \bibfield  {author} {\bibinfo {author} {\bibfnamefont {A.}~\bibnamefont
  {Le}}, \bibinfo {author} {\bibfnamefont {W.}~\bibnamefont {Daughton}},
  \bibinfo {author} {\bibfnamefont {L.-J.}\ \bibnamefont {Chen}}, \ and\
  \bibinfo {author} {\bibfnamefont {J.}~\bibnamefont {Egedal}},\ }\href
  {\doibase 10.1002/2017GL072522} {\bibfield  {journal} {\bibinfo  {journal}
  {Geophysical Research Letters}\ }\textbf {\bibinfo {volume} {44}},\ \bibinfo
  {pages} {2096} (\bibinfo {year} {2017})},\ \bibinfo {note}
  {2017GL072522}\BibitemShut {NoStop}%
\bibitem [{\citenamefont {Hesse}, \citenamefont {Zenitani},\ and\ \citenamefont
  {Klimas}(2008)}]{hesse:2008}%
  \BibitemOpen
  \bibfield  {author} {\bibinfo {author} {\bibfnamefont {M.}~\bibnamefont
  {Hesse}}, \bibinfo {author} {\bibfnamefont {S.}~\bibnamefont {Zenitani}}, \
  and\ \bibinfo {author} {\bibfnamefont {A.}~\bibnamefont {Klimas}},\ }\href
  {\doibase http://dx.doi.org/10.1063/1.3006341} {\bibfield  {journal}
  {\bibinfo  {journal} {Physics of Plasmas}\ }\textbf {\bibinfo {volume}
  {15}},\ \bibinfo {eid} {112102} (\bibinfo {year} {2008}),\
  http://dx.doi.org/10.1063/1.3006341}\BibitemShut {NoStop}%
\bibitem [{\citenamefont {Hesse}, \citenamefont {Kuznetsova},\ and\
  \citenamefont {Birn}(2004)}]{hesse:2004}%
  \BibitemOpen
  \bibfield  {author} {\bibinfo {author} {\bibfnamefont {M.}~\bibnamefont
  {Hesse}}, \bibinfo {author} {\bibfnamefont {M.}~\bibnamefont {Kuznetsova}}, \
  and\ \bibinfo {author} {\bibfnamefont {J.}~\bibnamefont {Birn}},\ }\href
  {\doibase 10.1063/1.1795991} {\bibfield  {journal} {\bibinfo  {journal}
  {Physics of Plasmas}\ }\textbf {\bibinfo {volume} {11}},\ \bibinfo {pages}
  {5387} (\bibinfo {year} {2004})},\ \Eprint
  {http://arxiv.org/abs/https://doi.org/10.1063/1.1795991}
  {https://doi.org/10.1063/1.1795991} \BibitemShut {NoStop}%
\bibitem [{\citenamefont {Ng}\ \emph {et~al.}(2015)\citenamefont {Ng},
  \citenamefont {Huang}, \citenamefont {Hakim}, \citenamefont {Bhattacharjee},
  \citenamefont {Stanier}, \citenamefont {Daughton}, \citenamefont {Wang},\
  and\ \citenamefont {Germaschewski}}]{ng:2015}%
  \BibitemOpen
  \bibfield  {author} {\bibinfo {author} {\bibfnamefont {J.}~\bibnamefont
  {Ng}}, \bibinfo {author} {\bibfnamefont {Y.-M.}\ \bibnamefont {Huang}},
  \bibinfo {author} {\bibfnamefont {A.}~\bibnamefont {Hakim}}, \bibinfo
  {author} {\bibfnamefont {A.}~\bibnamefont {Bhattacharjee}}, \bibinfo {author}
  {\bibfnamefont {A.}~\bibnamefont {Stanier}}, \bibinfo {author} {\bibfnamefont
  {W.}~\bibnamefont {Daughton}}, \bibinfo {author} {\bibfnamefont
  {L.}~\bibnamefont {Wang}}, \ and\ \bibinfo {author} {\bibfnamefont
  {K.}~\bibnamefont {Germaschewski}},\ }\href {\doibase
  http://dx.doi.org/10.1063/1.4935302} {\bibfield  {journal} {\bibinfo
  {journal} {Physics of Plasmas}\ }\textbf {\bibinfo {volume} {22}},\ \bibinfo
  {eid} {112104} (\bibinfo {year} {2015})}\BibitemShut {NoStop}%
\bibitem [{\citenamefont {{Allmann-Rahn}}, \citenamefont {{Trost}},\ and\
  \citenamefont {{Grauer}}(2018)}]{allmann:2018}%
  \BibitemOpen
  \bibfield  {author} {\bibinfo {author} {\bibfnamefont {F.}~\bibnamefont
  {{Allmann-Rahn}}}, \bibinfo {author} {\bibfnamefont {T.}~\bibnamefont
  {{Trost}}}, \ and\ \bibinfo {author} {\bibfnamefont {R.}~\bibnamefont
  {{Grauer}}},\ }\href@noop {} {\bibfield  {journal} {\bibinfo  {journal}
  {ArXiv e-prints}\ } (\bibinfo {year} {2018})},\ \Eprint
  {http://arxiv.org/abs/1801.07628} {arXiv:1801.07628 [physics.plasm-ph]}
  \BibitemShut {NoStop}%
\bibitem [{\citenamefont {Harris}(1962)}]{harris:1962}%
  \BibitemOpen
  \bibfield  {author} {\bibinfo {author} {\bibfnamefont {E.}~\bibnamefont
  {Harris}},\ }\href {http://dx.doi.org/10.1007/BF02733547} {\bibfield
  {journal} {\bibinfo  {journal} {Il Nuovo Cimento (1955-1965)}\ }\textbf
  {\bibinfo {volume} {23}},\ \bibinfo {pages} {115} (\bibinfo {year} {1962})},\
  \bibinfo {note} {10.1007/BF02733547}\BibitemShut {NoStop}%
\bibitem [{\citenamefont {Germaschewski}\ \emph {et~al.}(2016)\citenamefont
  {Germaschewski}, \citenamefont {Fox}, \citenamefont {Abbott}, \citenamefont
  {Ahmadi}, \citenamefont {Maynard}, \citenamefont {Wang}, \citenamefont
  {Ruhl},\ and\ \citenamefont {Bhattacharjee}}]{germaschewski:2016}%
  \BibitemOpen
  \bibfield  {author} {\bibinfo {author} {\bibfnamefont {K.}~\bibnamefont
  {Germaschewski}}, \bibinfo {author} {\bibfnamefont {W.}~\bibnamefont {Fox}},
  \bibinfo {author} {\bibfnamefont {S.}~\bibnamefont {Abbott}}, \bibinfo
  {author} {\bibfnamefont {N.}~\bibnamefont {Ahmadi}}, \bibinfo {author}
  {\bibfnamefont {K.}~\bibnamefont {Maynard}}, \bibinfo {author} {\bibfnamefont
  {L.}~\bibnamefont {Wang}}, \bibinfo {author} {\bibfnamefont {H.}~\bibnamefont
  {Ruhl}}, \ and\ \bibinfo {author} {\bibfnamefont {A.}~\bibnamefont
  {Bhattacharjee}},\ }\href {\doibase
  http://dx.doi.org/10.1016/j.jcp.2016.05.013} {\bibfield  {journal} {\bibinfo
  {journal} {Journal of Computational Physics}\ }\textbf {\bibinfo {volume}
  {318}},\ \bibinfo {pages} {305 } (\bibinfo {year} {2016})}\BibitemShut
  {NoStop}%
\bibitem [{\citenamefont {Birn}\ \emph {et~al.}(2001)\citenamefont {Birn},
  \citenamefont {Drake}, \citenamefont {Shay}, \citenamefont {Rogers},
  \citenamefont {Denton}, \citenamefont {Hesse}, \citenamefont {Kuznetsova},
  \citenamefont {Ma}, \citenamefont {Bhattacharjee}, \citenamefont {Otto},\
  and\ \citenamefont {Pritchett}}]{birn:2001}%
  \BibitemOpen
  \bibfield  {author} {\bibinfo {author} {\bibfnamefont {J.}~\bibnamefont
  {Birn}}, \bibinfo {author} {\bibfnamefont {J.~F.}\ \bibnamefont {Drake}},
  \bibinfo {author} {\bibfnamefont {M.~A.}\ \bibnamefont {Shay}}, \bibinfo
  {author} {\bibfnamefont {B.~N.}\ \bibnamefont {Rogers}}, \bibinfo {author}
  {\bibfnamefont {R.~E.}\ \bibnamefont {Denton}}, \bibinfo {author}
  {\bibfnamefont {M.}~\bibnamefont {Hesse}}, \bibinfo {author} {\bibfnamefont
  {M.}~\bibnamefont {Kuznetsova}}, \bibinfo {author} {\bibfnamefont {Z.~W.}\
  \bibnamefont {Ma}}, \bibinfo {author} {\bibfnamefont {A.}~\bibnamefont
  {Bhattacharjee}}, \bibinfo {author} {\bibfnamefont {A.}~\bibnamefont {Otto}},
  \ and\ \bibinfo {author} {\bibfnamefont {P.~L.}\ \bibnamefont {Pritchett}},\
  }\href {\doibase 10.1029/1999JA900449} {\bibfield  {journal} {\bibinfo
  {journal} {Journal of Geophysical Research: Space Physics}\ }\textbf
  {\bibinfo {volume} {106}},\ \bibinfo {pages} {3715} (\bibinfo {year}
  {2001})}\BibitemShut {NoStop}%
\bibitem [{\citenamefont {Ng}\ \emph {et~al.}(2011)\citenamefont {Ng},
  \citenamefont {Egedal}, \citenamefont {Le}, \citenamefont {Daughton},\ and\
  \citenamefont {Chen}}]{ng:2011}%
  \BibitemOpen
  \bibfield  {author} {\bibinfo {author} {\bibfnamefont {J.}~\bibnamefont
  {Ng}}, \bibinfo {author} {\bibfnamefont {J.}~\bibnamefont {Egedal}}, \bibinfo
  {author} {\bibfnamefont {A.}~\bibnamefont {Le}}, \bibinfo {author}
  {\bibfnamefont {W.}~\bibnamefont {Daughton}}, \ and\ \bibinfo {author}
  {\bibfnamefont {L.-J.}\ \bibnamefont {Chen}},\ }\href {\doibase
  10.1103/PhysRevLett.106.065002} {\bibfield  {journal} {\bibinfo  {journal}
  {Phys. Rev. Lett.}\ }\textbf {\bibinfo {volume} {106}},\ \bibinfo {pages}
  {065002} (\bibinfo {year} {2011})}\BibitemShut {NoStop}%
\bibitem [{\citenamefont {Gibbs}\ and\ \citenamefont {Su}(2002)}]{gibbs:2002}%
  \BibitemOpen
  \bibfield  {author} {\bibinfo {author} {\bibfnamefont {A.~L.}\ \bibnamefont
  {Gibbs}}\ and\ \bibinfo {author} {\bibfnamefont {F.~E.}\ \bibnamefont {Su}},\
  }\href {\doibase 10.1111/j.1751-5823.2002.tb00178.x} {\bibfield  {journal}
  {\bibinfo  {journal} {International Statistical Review}\ }\textbf {\bibinfo
  {volume} {70}},\ \bibinfo {pages} {419} (\bibinfo {year} {2002})}\BibitemShut
  {NoStop}%
\bibitem [{\citenamefont {Abdelmalik}\ and\ \citenamefont {van
  Brummelen}(2016)}]{abdelmalik:2016}%
  \BibitemOpen
  \bibfield  {author} {\bibinfo {author} {\bibfnamefont {M.~R.~A.}\
  \bibnamefont {Abdelmalik}}\ and\ \bibinfo {author} {\bibfnamefont {E.~H.}\
  \bibnamefont {van Brummelen}},\ }\href {\doibase 10.1007/s10955-016-1529-5}
  {\bibfield  {journal} {\bibinfo  {journal} {Journal of Statistical Physics}\
  }\textbf {\bibinfo {volume} {164}},\ \bibinfo {pages} {77} (\bibinfo {year}
  {2016})}\BibitemShut {NoStop}%
\bibitem [{\citenamefont {Hesse}, \citenamefont {Winske},\ and\ \citenamefont
  {Kuznetsova}(1995)}]{hesse:1995}%
  \BibitemOpen
  \bibfield  {author} {\bibinfo {author} {\bibfnamefont {M.}~\bibnamefont
  {Hesse}}, \bibinfo {author} {\bibfnamefont {D.}~\bibnamefont {Winske}}, \
  and\ \bibinfo {author} {\bibfnamefont {M.~M.}\ \bibnamefont {Kuznetsova}},\
  }\href {\doibase 10.1029/95JA01559} {\bibfield  {journal} {\bibinfo
  {journal} {Journal of Geophysical Research: Space Physics}\ }\textbf
  {\bibinfo {volume} {100}},\ \bibinfo {pages} {21815} (\bibinfo {year}
  {1995})}\BibitemShut {NoStop}%
\end{thebibliography}%

\end{document}